\documentclass[%
 reprint,
 amsmath,amssymb,
prb,
]{revtex4-2}

\usepackage{graphicx}
\usepackage{dcolumn}
\usepackage{bm}
\usepackage{color}
\usepackage{ulem}

\begin{document}

\preprint{APS/123-QED}

\title{First-principles study of ultrafast and nonlinear optical properties \\ of graphite thin films}

\author{Mitsuharu Uemoto}
 \affiliation{Department of Electrical and Electronic Engineering, Kobe University, Kobe 657-8501, Japan}
 \email{uemoto@eedept.kobe-u.ac.jp}
\author{Shintaro Kurata}%
\affiliation{Physical Engineering Group, Technology Platform Center, Technology \& Intelligence Integration, IHI Corporation, Yokohama 235-8501, Japan}
\author{Norihito Kawaguchi}
\affiliation{Collaboration \& Marketing Group, Technology Planning Department, Technology \& Intelligence Integration, IHI Corporation, Yokohama 235-8501, Japan}
\author{Kazuhiro Yabana}
\affiliation{Center for Computational Sciences, University of Tsukuba, 1-1-1 Tennodai, Tsukuba, 305-8577, Japan}

\date{\today}

\begin{abstract}

We theoretically investigate ultrafast and nonlinear optical properties of graphite thin films
based on first-principles time-dependent density functional theory.
We first calculate electron dynamics in a unit cell of graphite under a strong pulsed electric field
and explore the transient optical properties of graphite.
It is shown that the optical response of graphite shows a sudden change from conducting to
insulating phase at a certain intensity range of the applied electric field. 
It also appears as a saturable absorption, the saturation in the energy transfer from the electric field to electrons.
We next investigate a light propagation in graphite thin films by solving coupled dynamics of
the electrons and the electromagnetic fields simultaneously.
It is observed that the saturable absorption manifests in the propagation 
with small attenuation in the spatial region where the electric field amplitude is about $4 \sim 7 \times 10^{-2}$ V/\AA.

\end{abstract}
\maketitle

\section{Introduction}

Owing to rapid progresses of laser technologies, it has become possible to utilize
a pulsed light freely selecting its intensity and duration \cite{Brabec2000rmp}. 
As for the duration of laser pulses, it has become possible to produce light pulses
as short as a few tens of attosecond \cite{Krausz2009rmp}. 
Using intense and ultrashort laser pulses, there have been observed a number of 
intriguing phenomena in solids.
For example, high harmonic generation from solids has been extensively explored
in the last decade \cite{Ghimire2011}.
An induced electric current in glass by the ultrashort pulsed light has been observed,
indicating that the insulator can be changed into a conducting material 
in a very short time-scale \cite{Schiffrin2013current}.
Using further intense laser pulses, nonthermal laser processing is expected as an
efficient means of micro-fabrication of materials \cite{Chichkov1996}.

The present paper aims to report a theoretical and computational analysis on the ultrafast 
and nonlinear optical responses of graphite thin films.
Graphite is a semi-metallic layered material of carbon atoms arranged in a honeycomb lattice. 
Monolayered carbon structures such as graphene and single-wall carbon nanotube have 
been attracting enormous attention for their unusual optical properteis \cite{Bonaccorso2010, Yu2017}.
For example, large optical nonlinearities such as harmonic generation \cite{Yoshikawa2017}, 
Kerr effects \cite{Cheng2015a, chu2012ultrafast} have been reported.
In particular, ultrafast saturable absorption has been extensively investigated and
utilized as satulable absorber in a mode-locked fiber laser 
\cite{Bao2009, Kumar2009, Sun2010, Xing2010,  Norris2012, Omi2015, Sobon2015, Wang2016ACS, Demetriou2016, Kurata2018}. 
Using further intense laser pulses, laser processing of carbon materials has also attracted 
attention \cite{Choi1999,Janulewicz2014}.
Since it has been known that micro-fabrication of some carbon materials are difficult by
ordinary methods \cite{Weber2011cfrp}, efficient processing using ultrashort pulsed light is highly expected.

To obtain reliable understanding for the ultrafast and nonlinear optical properties, 
it is essential to describe microscopic electron motion under an optical field.
Optical properties of graphenes are characterized by the presence of the Dirac cone in the energy band. 
Although the valence and the conduction bands slightly overlap in the graphite making it semi-metal,
optical properties of graphite are also characterized mostly by the Dirac cone like structure.
There have been reported a number of theoretical analyses treating motion of electrons in the Dirac cone. 
They include analytical modeling and analysis \cite{Yu2017model,Mikhailov2011,Marini2017},
quasi-classical kinetic theory \cite{Mikhailov2017},
time-dependent solution for the optical Bloch equation \cite{Ishikawa2010,Chizhova2017},
and time-dependent solution for the density matrix equation \cite{Malic2011,Zhang2011,Cheng2015a}.
There have been carried out first-principles calculations including all valence electrons
for nonlinear optical properties in frequency domain \cite{Guo2004,Wang2016}.
Descriptions of nonlinear light propagation is another important issue to understand optical
responses of multilayer graphenes and graphite thin films, and also laser processing of carbon materials. 
There have been used a transfer matrix method \cite{Dean2010}. Nonlinear propagation method 
was also developed \cite{Castello2020}.

In the present study, we will employ a first-principles computational approach based on time-dependent 
density functional theory (TDDFT) \cite{runge1984density,Ullrich2012}.
The TDDFT has been known as a useful method to investigate optical properties of molecules
and solids with a reasonable accuracy for a moderate computational cost.
Solving the time-dependent Kohn-Sham (TDKS) equation in time domain \cite{yabana1996time,Bertsch2000}, 
it is possible to describe ultrafast and nonlinear electronic dynamics in matters.
Applications of such approaches include nonlinear optical constants \cite{uemoto2019nonlinear},
high harmonic generations \cite{Otobe2016,Floss2018,Breton2018},  
light-matter energy transfer \cite{Sato2015, yamada2018time}.

Solving the TDKS equation in time domain, it is possible to investigate ultrafast electron 
dynamics in graphite induced by a pulsed electric field without any perturbative approximations.
We can describe dynamics of whole $sp$ valence electrons including the dynamics at the Dirac cone.
Using ordinary functionals, however, collisional effects cannot be included sufficiently,
although carrier relaxation dynamics is known to be important in graphite \cite{Breusing2009}
and graphene \cite{breusing2011ultrafast}.
This limits the validity of the TDDFT approach to a short duration before the relaxation becomes
significant. The relaxation time should depend on the laser parameters and is considered about
a few tens of femtosecond \cite{Breusing2009, breusing2011ultrafast, Marini2017,Cheng2015a}.

A theoretical description of the propagation of light is another important subject.
There has been a progress to develop combined simulations of TDDFT for electron dynamics 
and electromagnetism analysis for a light propagation adopting a multiscale strategy \cite{yabana2012time}.
The method has been successfully applied to analyze ultrafast and nonlinear light propagations:
the attosecond transient absorption spectroscopy mimicking pump-probe experiments \cite{Lucchini2016}
and  the spatial distribution of energy deposition that is the basic information to analyze laser
processing \cite{Sato2015-2, Sommer2016}.

This paper is organized as follows: 
In Sec.~\ref{sec_theory}, we provide a formalism and a computational method based on TDDFT.
Electron dynamics in a unit cell of graphite induced by pulsed electric fields are discussed in Sec.~\ref{sec_unitcell}.
Light propagations through graphite thin films are discussed in Sec.~\ref{sec_propagation}. 
Finally, in Sec.~\ref{sec_summary}, a summary will be presented.

\section{Theory}
\label{sec_theory}

\subsection{Electron dynamics in a unit cell}

We describe a time evolution of electron orbitals in a unit cell of graphite under a strong pulsed electric field 
by solving the TDKS equation for Bloch orbitals.
Expressing the applied electric field using a vector potential $\mathbf{A}(t)$ in the velocity 
gauge \cite{Bertsch2000}, we have:
\begin{align}
  i \frac{\partial}{\partial t}
  u_{b \mathbf{k}}(\mathbf{r}; t)
  =&
  H[\mathbf{A}]
  u_{b \mathbf{k}}(\mathbf{r}; t)
  \;,
  \label{eq_tdks}
\end{align}
where $u_{b \mathbf{k}}$ is the Bloch orbitals with band index $b$ and wavenumber $\mathbf{k}$. 
The Hamiltonian $H[\mathbf{A}]$ is given by
\begin{align}
 H[\mathbf{A}]
  =&
    \frac{1}{2m} \left[
      \hat{\mathbf{p}}
      +
      \hbar \mathbf{k}
      +
      \frac{\mathbf{A}(t)}{c}
    \right]^2
  \notag \\
  &
    +
    V_{\mathrm{ps}}(\mathbf{r})
    +
    V_{\mathrm{H}}[n](\mathbf{r})
    +
    V_{\mathrm{xc}}[n](\mathbf{r})
  \;,
  \label{eq_tdks_h}
\end{align}
where 
$V_{\mathrm{ps}}$, $V_{\mathrm{H}}$ and $V_{\mathrm{xc}}$ are ionic pseudopotentials, Hartree and 
exchange-correlation potentials, respectively.
The Hartree and the exchange-correlation potentials depend on the electron density $n(\mathbf{r}; t)$ 
that is expressed as: 
\begin{align}
  n(\mathbf{r}; t)
  =&
  \frac{2}{\Omega_\mathrm{BZ}}
  \int_{\Omega_\mathrm{BZ}}
  \mathrm{d}\mathbf{k'}
  \sum_{b}^\mathrm{(occ)}
  \vert
  u_{b \mathbf{k}'}(\mathbf{r}; t)
  \vert^2
  \;,
  \label{eq_density_exact}
\end{align}
where $\Omega_{\mathrm{BZ}}$ is the volume of the Brillouin zone (BZ). 
The summation of $b$ is taken over all occupied bands.
The integration in the BZ is carried out by numerical quadrature using appropriately selected sampling points.
Therefore, Eq~(\ref{eq_density_exact}) can be rewritten as
\begin{align}
  n(\mathbf{r}; t)
  =&
  2
  \sum_{\mathbf{k}_i}^{BZ}
  \sum_{b}^\mathrm{(occ)}
  w_i
  \vert
  u_{b \mathbf{k}_i}(\mathbf{r}; t)
  \vert^2
  \;,
  \label{eq_density}
\end{align}
where $\mathbf{k}_i$ and $w_i$ are the sampled $k$-points and their weighting coefficients, respectively.
A uniformly distributed $\mathbf{k}_i$ with equal $w_i$ are often used for calculations of periodic crystals.
In this study, however, we use a non-uniform $k$-point sampling in the BZ to obtain convergent results 
with less computational costs. Details will be explained in Sec. \ref{sec_model}. 

From the Bloch orbitals, the electric current density averaged over the unit cell, $\mathbf{J}(t)$, is given as below:
\begin{align}
  \mathbf{J}[{\mathbf A}](t)
  = &
  -
  \frac{2}{\Omega_\mathrm{cell}} \sum_{b \mathbf{k}_i} w_i
   \int_{\Omega_\mathrm{cell}} \mathrm{d} \mathbf{r}' \;
  u_{b \mathbf{k}_i}^\star(\mathbf{r}', t)
  \notag \\ &
  \Biggl(
    \hat{\mathbf{p}} + \mathbf{k}
    +
    \frac{\mathbf{A}(t)}{c}
    -i
    \bigl[
      \hat{\mathbf{r}},
      V_\mathrm{NL}
    \bigr]
  \Biggr)
  u_{b \mathbf{k}_i}(\mathbf{r}', t)
  \; ,
  \label{eq_current}
\end{align}
where $V_{\mathrm{NL}}$ is the nonlocal part of the pseudopotential $V_{\mathrm{ps}}$, and $\Omega_\mathrm{cell}$ is the volume of the unit cell.

\subsection{Light propagation}
\label{sec:multiscale}

We next consider a description of light propagation in a thin film of graphite.
We will use "Maxwell+TDDFT multiscale method" \cite{yabana2012time}, calculating light propagations in solids 
by combining electromagnetics field analysis and electronic dynamics calculations.
We note the the spatial scale of the electromagnetic fields of the propagating light is a few hundreds of nanometers, 
and is much larger than that of the microscopic electronic dynamics.
In order to overcome the mismatch of the two spatial scales, we utilize a multiscale method 
introducing the two coordinate systems for the macroscopic and microscopic dynamics. 
\color{red}

The multiscale method is a natural extension of a macroscopic electromagnetism
that is usually used to describe light propagation in a medium.
In the macroscopic electromagnetism, one usually solve macroscopic Maxwell equation
using a grid system in which the grid spacing is typically much larger than atomic scale.
The microscopic dynamics is taken into account through a constitutive relation
that relates, for example, the electric field and the electric current locally, 
$J(t) = J[E(t)]$.
To describe a propagation of an intense pulsed light, we cannot use any perturbative 
constitutive relations.
Instead, we solve the TDKS equation to relate the current density and the
electric field. It is noted that we need to consider electron dynamics at each grid
point that is used to solve the macroscopic Maxwell equation, since electron dynamics
at each point is different from each other.
The microscopic coordinate is used to solve the TDKS equation to calculate 
microscopic electron dynamics.
\color{black}

For the macroscopic electromagnetic fields, we solve the one-dimensional Maxwell's equation:
\begin{align}
  \frac{1}{c^2}
  \frac{\partial^2 A_X(t)}{\partial t^2}
  -
  \frac{\partial^2 A_X(t)}{\partial X^2}
  =&
  \frac{4 \pi}{c} J_X(t)
  \;,
  \label{eq_maxwell1d}
\end{align}
where $X$ is the macroscopic coordinate variable, $A_X(t)$ is the vector potential field at $X$.
At each point $X$, we consider a microscopic electronic system.
At the microscopic scale, we consider an electron dynamics of infinitely extended system
under a spatially  uniform electric field specified by $A_X(t)$ where we treat $X$ as a parameter.
The electron motion is described by Bloch orbitals, $u_{X b \mathbf{k}}$,
that satisfy
\begin{align}
  i \frac{\partial}{\partial t}
  u_{Xb \mathbf{k}}(\mathbf{r}; t)
  =&
  H[A_X]
  u_{Xb \mathbf{k}}(\mathbf{r}; t)
  \;,
  \label{eq_TDKSmultiscale}
\end{align}
The current density $J_X(t)$ is determined by Eq.~(\ref{eq_current}) with
$J_X(t) = J[A_X](t)$.
We solve the TDKS and the Maxwell equations simultaneously exchanging $A_X(t)$ and  $J_X(t)$ 
at every time step. 

\color{red}
It should be noted that the present multiscale formalism returns to the ordinary
macroscopic electromagnetism when a field is sufficiently weak.
\color{black}
The detail of our formalism is explained in the reference \cite{yabana2012time}.

\subsection{Computational details}

\begin{figure}
  \resizebox{0.5\textwidth}{!}{%
    \includegraphics{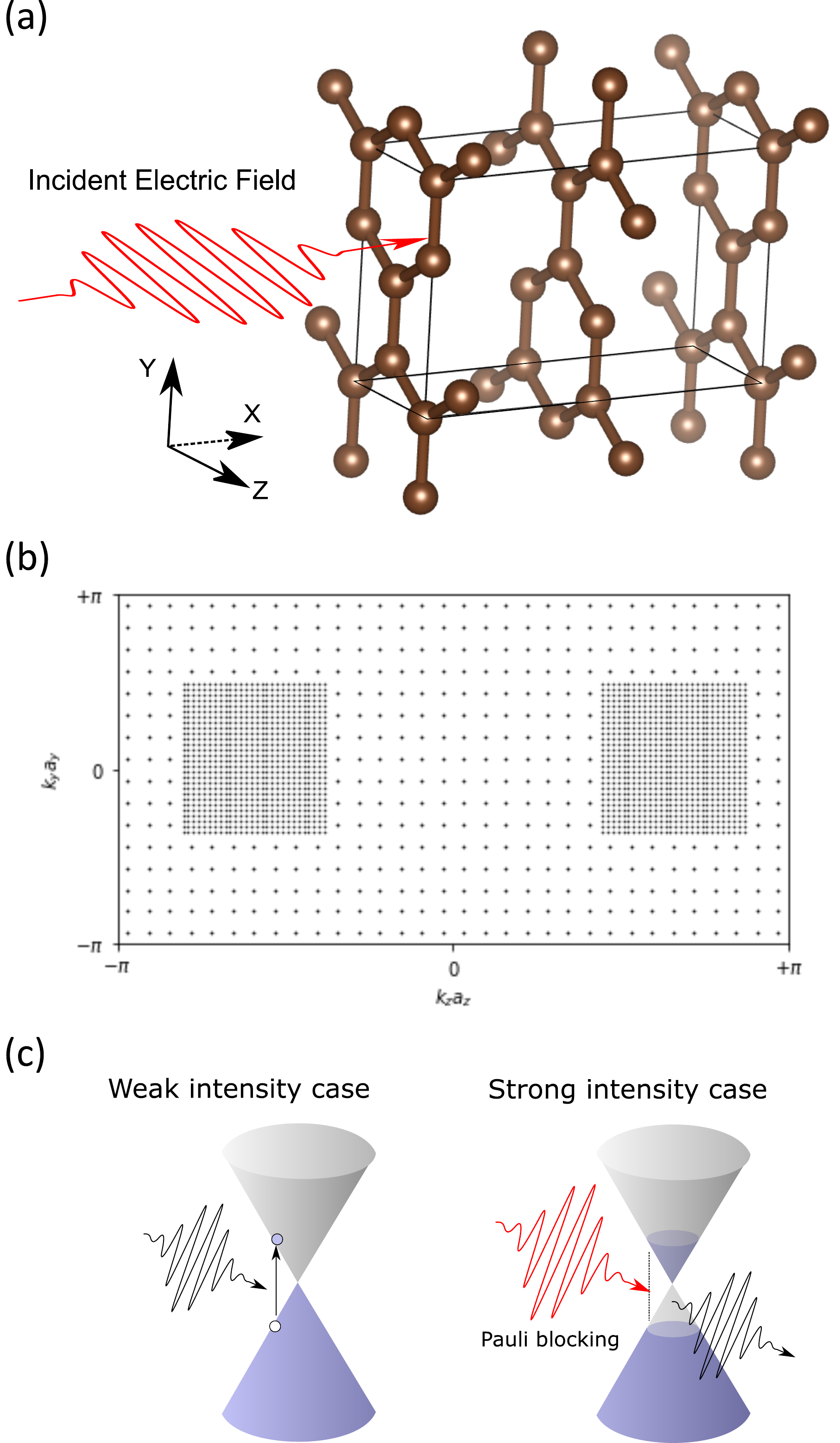}
  }
  \caption{
  Schematic illustration of a calculated system:
  (a) Linearly polarized electric filed irradiates normaly on an ABA-stacked graphite crystal 
  for which we use a rectangular unit cell with 8 carbon atoms.
  (b) Non-uniform $k$-point distribution to sample the entire BZ that contains 1,656 coarse points and 6,272 dense points.
  \color{red}
  (c) Band structure and electron occupation in the weak (left) and strong (right) field cases.
  }
  \label{fig_structure}
\end{figure}

\label{sec_model}
We consider an ABA-stacked graphite crystal. Fig.~\ref{fig_structure}(a) illustrates the crystal structure.
The lattice constant and the interlayer distance are set as $1.42$ and $3.35$~ \AA, respectively.
We use a rectangular unit cell of $6.70 \times 4.25 \times 2.46$~\AA, which contains $8$ carbon atoms.
For electron-ion interaction, we employ norm-conserving pseudopotential \cite{troullier1991efficient} 
having $4$ valence electrons in a single atom.
For the exchange-correlation potential, the adiabatic local density approximation (ALDA) with Perdew-Zunger 
functional \cite{perdew1981self} is adopted. 
To express Bloch orbitals, we use a three-dimensional Cartesian grid representation with a finite-difference scheme for differentiation operators.
 The crystalline unit cell is divided into $26 \times 16 \times 16$ uniform grids. The resulting grid spacing is about $0.26 \times 0.27 \times0.15$~\AA; 
the grid spacing for $z$-direction ($E$-field direction) is set to be finer than the other spatial axes.
For the summation over $k$-space in Eqs.~(\ref{eq_density})-(\ref{eq_current}), the entire BZ is sampled by $7,928$ non-uniformly generated $k$-points \color{red} [see  Fig.~\ref{fig_structure}(b)]. 
\color{black} Since electronic excitations dominate around the Dirac cone region in the $k$-space 
\color{red} [see Fig.~\ref{fig_structure}(c)] \color{black}, we use a non-uniform sampling, 
a dense sampling for focused regions and a corse sampling in the other areas, to improve the accuracy while avoiding 
the increase of the computational cost. 
In the $ k_x $ (interlayer) direction, the BZ is divided into four. Each $k_y k_z$-layer is sampled by 414 course points 
and 1,568 dense points as shown in Fig.~\ref{fig_structure}(b).
The weighting coefficients are about $w_i \approx 4.88 \times 10^{-4}$ and $3.05\times10^{-5}$, respectively. 
The time evolution is calculated using an enforced time reversal symmetry (ETRS) \cite{castro2004propagators} propagator with a time step of  ${\Delta}t=0.04~\mathrm{au} (0.96~\mathrm{as})$.
For computation, we use an open-source TDDFT program package, SALMON
(Scalable Ab-initio Light-Matter simulator for Optics and Nanoscience)
which has been developed in our group \cite{noda2018salmon}.

\section{Electron dynamics in a unit cell}
\label{sec_unitcell}

\subsection{Dielectric function}

Before discussing nonlinear and ultrafast responses, we first show calculation of the dielectric function of graphite
to confirm the reliability of our model in the linear response.
We utilize a real-time scheme to calculate the dielectric function as described below \cite{yabana2012time}. 
Here, as a vector potential in Eqs.~(\ref{eq_tdks_h}) and (\ref{eq_current}), we adopt a Heaviside step function with a small amplitude $\delta A$:
\begin{align}
  A_z(t) &= \delta{A}  \; \theta(t)
  \;,
\end{align}
which corresponds to an impulsive electric field described by the Dirac delta function given as $\mathbf{E}(t) = - \partial_t {\mathbf{A}}(t)/c$.
For this perturbation, we calculate the induced current density~(\ref{eq_current}) and take the Fourier transformation to obtain the
conductivity and the dielectric constant as a function of frequency $\omega$:
\begin{align}
  \sigma_{zz}(\omega)
  =&
  \frac{
    \int \mathrm{d}t \; J_z(t) e^{i\omega t}
  }{
    \int \mathrm{d}t \; E_z(t) e^{i\omega t}
  }
\end{align}
and
\begin{align}
  \epsilon_{zz}(\omega)
  =&
  1 + \frac{4 \pi i}{\omega} \sigma_{zz}(\omega)
  \;.
\end{align}

\begin{figure}
  \resizebox{0.5\textwidth}{!}{%
    \includegraphics{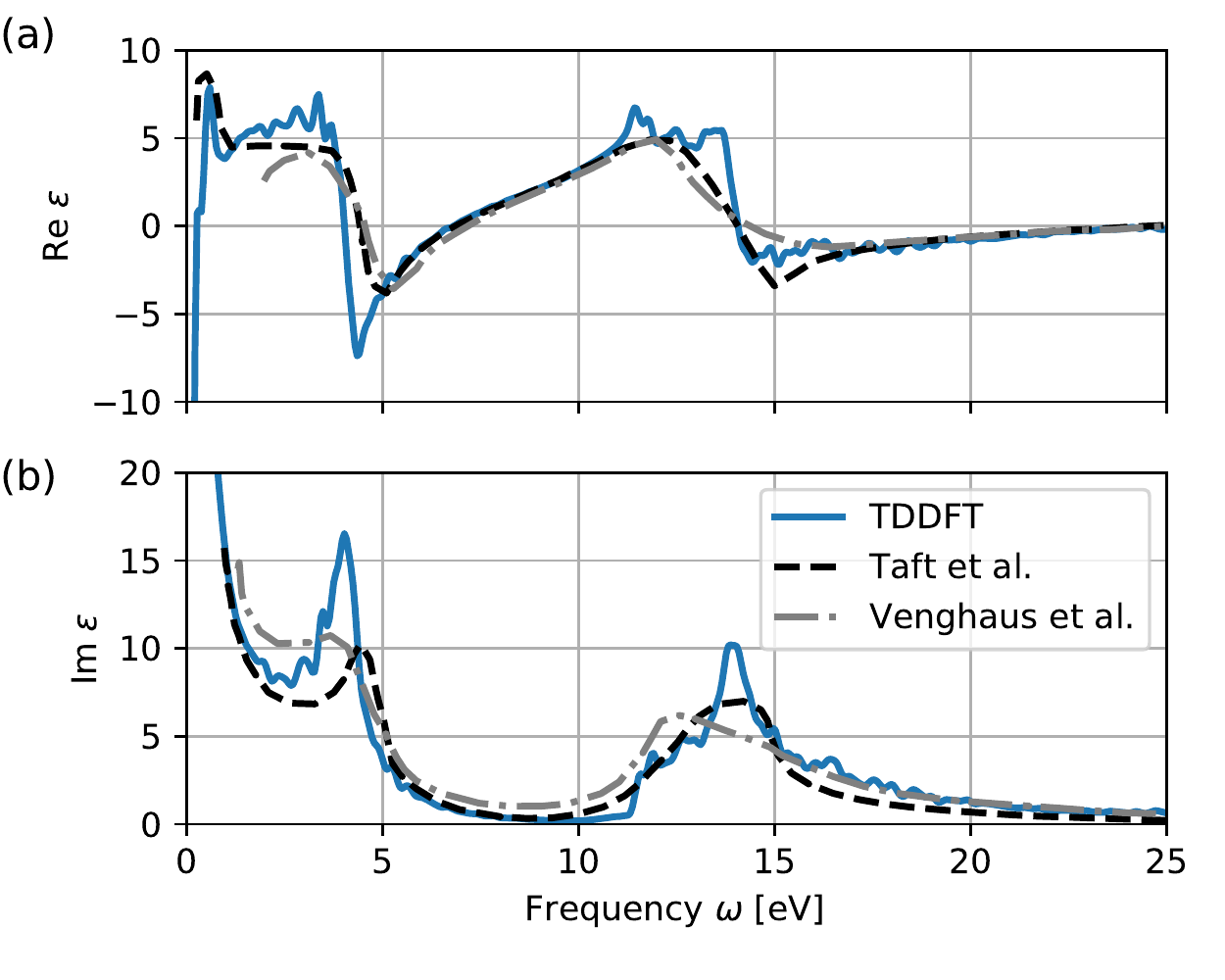}
  }
  \caption{
  Dielectric function of a crystalline graphite in the direction parallel to the layer : (a) real part $\mathrm{Re} \varepsilon_{zz}$ and (b) imaginary part $\mathrm{Im} \varepsilon_{zz}$. 
  The experimental spectra are also plotted as black \cite{Taft1965} and gray \cite{venghaus1975redetermination} broken curves.
  }
  \label{fig_epsilon}       
\end{figure}

In Fig.~\ref{fig_epsilon}, we show the calculated dielectric function $\epsilon_{zz}(\omega)$ (blue solid curve).
For comparison, we also show the experimental spectra  \cite{Taft1965, venghaus1975redetermination} (broken curves).
Since we include all valence orbitals, calculated dielectric function shows reasonable agreement with measurements
for wide energy region. 
In later sections, we will mostly discuss the interaction of laser pulses with a central frequency of 
$\omega_1 \sim 1.55~\mathrm{eV}$.
At this frequency, 
the dielectric constant is $\epsilon_\mathrm{TDDFT}(\omega_1)=5.3+10i$ in our TDDFT calculation.
This is close to the measured value, $\epsilon_\mathrm{exp}(\omega_1) \approx 5+9i$ \cite{Taft1965}.

\subsection{Response to pulsed electric fields}

Next, we investigate electronic dynamics and optical responses of graphite induced by 
a short and strong pulsed electric field.
We will use the following waveform:
\begin{align}
  A_z(t)
  =&
  \frac{E_\mathrm{max}}{c \omega_1} f(t) \cos \omega_1 t
  \label{eq_pulse}
\end{align}
with $\cos^2$-type envelope function
\begin{align}
  f(t)
  =&
  \begin{cases}
  \left[
    \cos \left(\pi (t - T_P / 2) / T_P \right)
  \right]^2
  &
  0 \leq t \leq T_P
  \\
  0
  &
  \mathrm{otherwise}
  \end{cases}
  \;,
  \label{eq_cos2}
\end{align}
where $E_\mathrm{max}$ is the maximum amplitude of the electric field and $T_P$ is the duration of the pulse envelope.
The pulse duration is conventionally  expressed using the full-width half-maximal (FWHM); 
for the $\cos^2$-shaped envelope function
adopted here, the FWHM duration is given by  $T_{\mathrm{FWHM}} \approx 0.364T_P$.

\begin{figure}
  \resizebox{0.45\textwidth}{!}{%
    \includegraphics{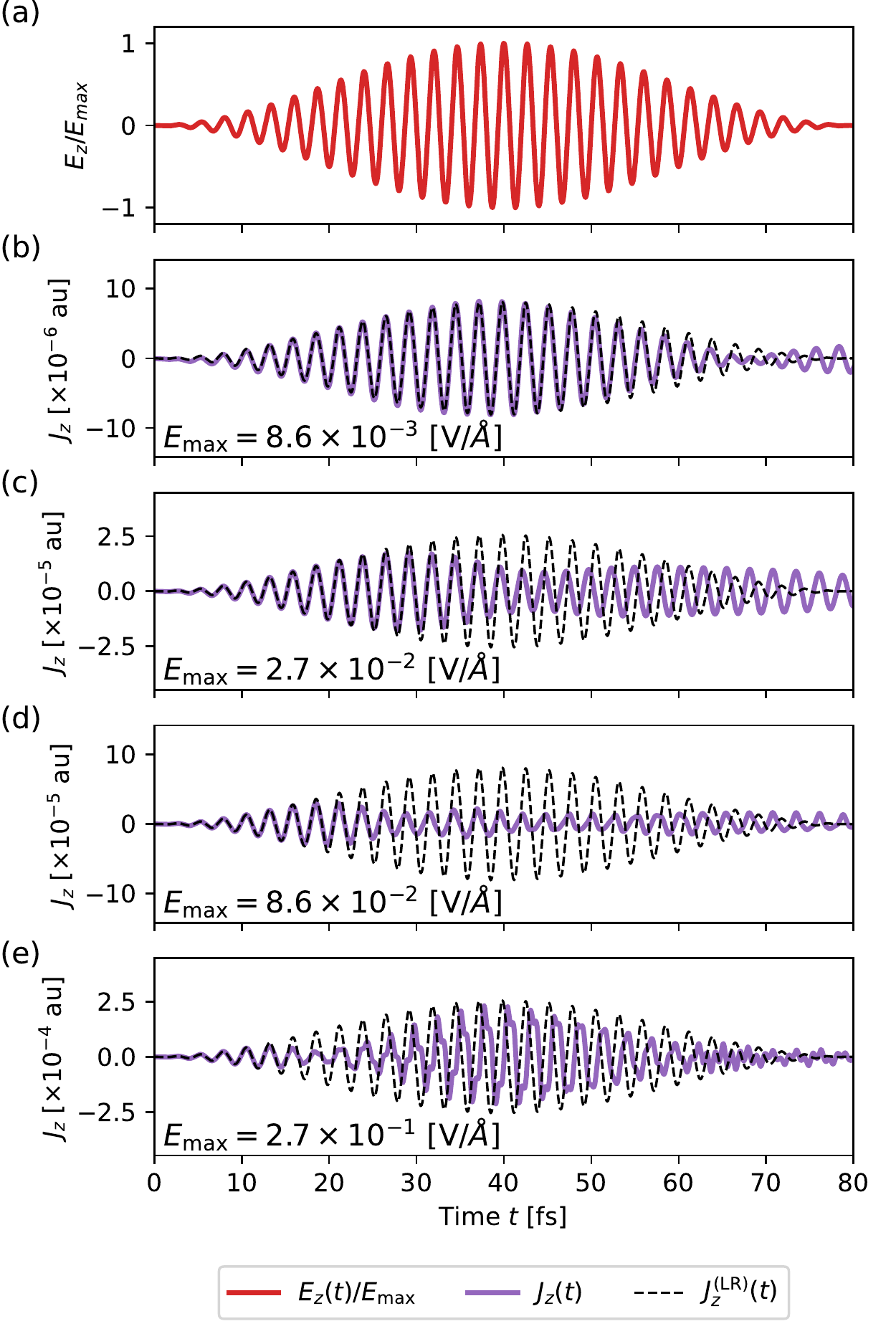}
  }
  \caption{
  Applied electric field (a) and induced current density (b-e) for four different maximal field amplitudes $E_\mathrm{max}$.
   The central frequency and the pulse duration are set to $\omega_1=1.55~\mathrm{eV}$ and 
   $T_P=80~\mathrm{fs}$, respectively.
  In $(b-e)$, purple lines show calculated current density using Eq.~(\ref{eq_current}).
  Black broken-line shows the linear results with $\sigma = 0.047 - 0.018 i ~\mathrm{au}$.
  }
  \label{fig_current}       
  \end{figure}

  Figure~\ref{fig_current} shows (a) temporal profiles of the applied pulsed electric field and 
  (b-e) induced currents for four different amplitudes of the electric field: 
  $E_\mathrm{max} = 8.7 \times 10^{-3} \sim 2.7 \times 10^{-1}$~V/\AA.
  The pulse duration is taken to be $T_P=80$~fs ($T_\mathrm{FWHM} \approx 30$~fs).
  The purple solid curve shows the current density calculated by Eq.~(\ref{eq_current}), 
  and the dotted black curve shows the current density assuming a linear response,
  using a conductivity: 
  $J_z^{(LR)}(t) = [{\rm Re} \sigma(\omega_1)] E_z(t) - [{\rm Im} \sigma(\omega_1)/\omega_1] \dot{E}_z(t)$
  with 
  the conductivity from the TDDFT calculation,
  $\sigma =0.047 - 0.018i~\mathrm{au}$.
  
  From the calculation, we find the following nonlinear behavior as the amplitude $E_\mathrm{max}$ increases.
  When the applied field amplitude is weak, $J_z(t)$ agrees well with the estimate assuming the linear response, $J_z^{(LR)}(t)$,
  as seen in (b) at $E_\mathrm{max} = 8.6 \times 10^{-3}$~V/\AA.
  Here the optical response is conducting: the current is mostly in phase with the electric field, $E(t)$.
  As the field amplitude increases, the current starts to depart from the linear response.
  At $E_\mathrm{max} = 2.7 \times 10^{-2}$~V/\AA~ that is shown in (c),
  the current gradually attenuates during the pulse irradiation 
  and there also appears a phase change at around $t = 40$ fs.
  The attenuation becomes maximum in (d) where the field strength is $E_\mathrm{max} = 8.6 \times 10^{-2}$~V/\AA.
  Further increasing the field strength, the amplitude of the current $J_z(t)$ increases again in (e) at the
  field amplitude of $2.7 \times 10^{-1}$~V/\AA. Here an appearance of high-frequency oscillations
  is also observed.
  
  We consider that the suppression of the induced current is caused by the saturable absorption (SA).
  To understand the temporal change of the optical response, it is useful to analyze the relative phase shift
  between the applied electric field and the induced current, as well as the amplitude.
  The induced current is in phase with the electric field for conducting case, $J \approx \sigma E$ (conducting phase),
  while the induced current has a phase difference of $\pi / 2 $ to the electric field (insulating phase)
  for insulators, since the current is expressed as the time derivative of the polarization, 
  $J = \partial t P(t) = \chi \partial_t E$.
  
  To analyze the phase change quantitatively, we show the short-time Fourier transform:
  \begin{align}
  \sigma(\omega; t)=
  \frac{
    \int J_z(t') e^{i \omega t'} w_b(t'-t) \; \mathrm{d}t'
  }{
    \int E_z(t') e^{i \omega t'} w_b(t'-t) \; \mathrm{d}t'
  },
  \label{eq_stft}
\end{align}
where $w_b(t)$ is the Blackman's window function with the width of $10~\mathrm{fs}$.
\begin{figure}
  \resizebox{0.5\textwidth}{!}{%
    \includegraphics{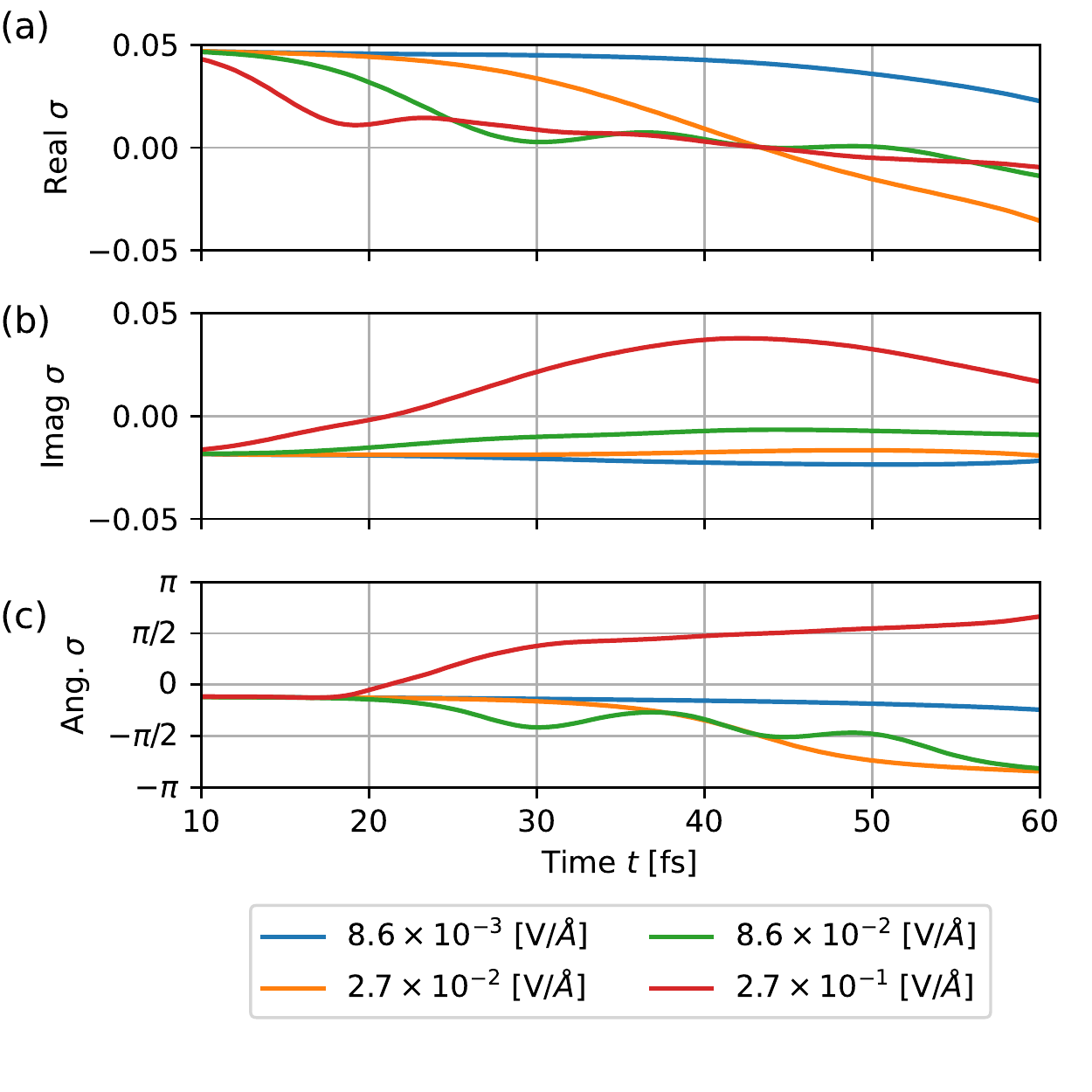}
  }
  \caption{ 
  Temporal-evolution of the complex conductivity calculated from the short-time Fourier transform (\ref{eq_stft}). 
  The real part (a), imaginary part (b), and phase-angle (c) of the conductivity are plotted for three cases of
  maximum electric field amplitude.
   }
  \label{fig_phase}       
  \end{figure}
Figure~\ref{fig_phase} shows the temporal evolution of the real and imaginary parts, and the phase of the
conductivity. 
At small $E_\mathrm{max}$, the conductivity behaves as a constant in time, indicating that
the response can be described by the linear response theory.
\color{red}{\sout{
A rather sudden change at $t>60$~fs is caused by an artifact of the short time Fourier transform.}} \color{black}
As $E_\mathrm{max}$ increases, the reduction of the real part $\mathrm{Re}~\sigma(\omega;t)$  
and the progression of the phase $\mathrm{ang}\;\sigma(\omega;t)$ arises. 
At $E_\mathrm{max} = 8.6 \times 10^{-2}$~V/\AA, the phase-difference goes to $\pi/2$ 
at the central time of the pulse.
These results indicate that the optical property of graphite changes from conductor to insulator under 
the intense optical field. 
\color{red}
The red curve in Fig.~\ref{fig_phase} shows a transient conductivity under a strong
applied field corresponding to Fig.~\ref{fig_current}(e).
We find out a characteristic phase inversion in the imaginary part of the conductivity.
This change is considered to come from the Drude-like response of metallic media
due to an increase of excited carriers caused by the strong applied field.
We will later discuss the occupation distribution of this case.

\color{black}

Next, we investigate the electronic excitation energy, that is, the energy transfer from the applied electric
field to electrons in the unit cell.
From the induced current $\mathbf{J}(t)$ under the applied electric field $\mathbf{E}(t)$, 
the energy deposition per unit time and volume can be evaluated by $W(t)=\mathbf{E}(t) \cdot \mathbf{J}(t)$.
Therefore, we introduce the electronic excitation energy per atom at time $t$ by
\begin{align}
  E_{\mathrm{ex}}(t)
  =&
  \frac{
    \Omega_\mathrm{cell}
  }{
    N_\mathrm{atom}
  }
  \int_{-\infty}^t{
    \mathrm{d}t' \;
    \mathbf{E}(t) \cdot \mathbf{J}(t)
  }
  \;,
\end{align}
where $N_\mathrm{atom}$ denotes the number of atoms contained in the unit cell.
For a weak field, the excitation energy reduces
\begin{align}
  E^{\mathrm{(LR)}}_{\mathrm{ex}}(t)
  =&
  \frac{\Omega_\mathrm{cell}}{N_{\mathrm{atom}}}
  \int^t_{-\infty} {
    \mathrm{d}t' \;
    \mathrm{Re} \; \sigma_{zz}(\omega_1)
    \left[
      E_z(t')
    \right]^2
  }
  \;,
  \label{eq_excitation_linear}
\end{align}
if the frequency dependence of the conductivity is small.

\begin{figure*}
  \resizebox{1.00\textwidth}{!}{%
    \includegraphics{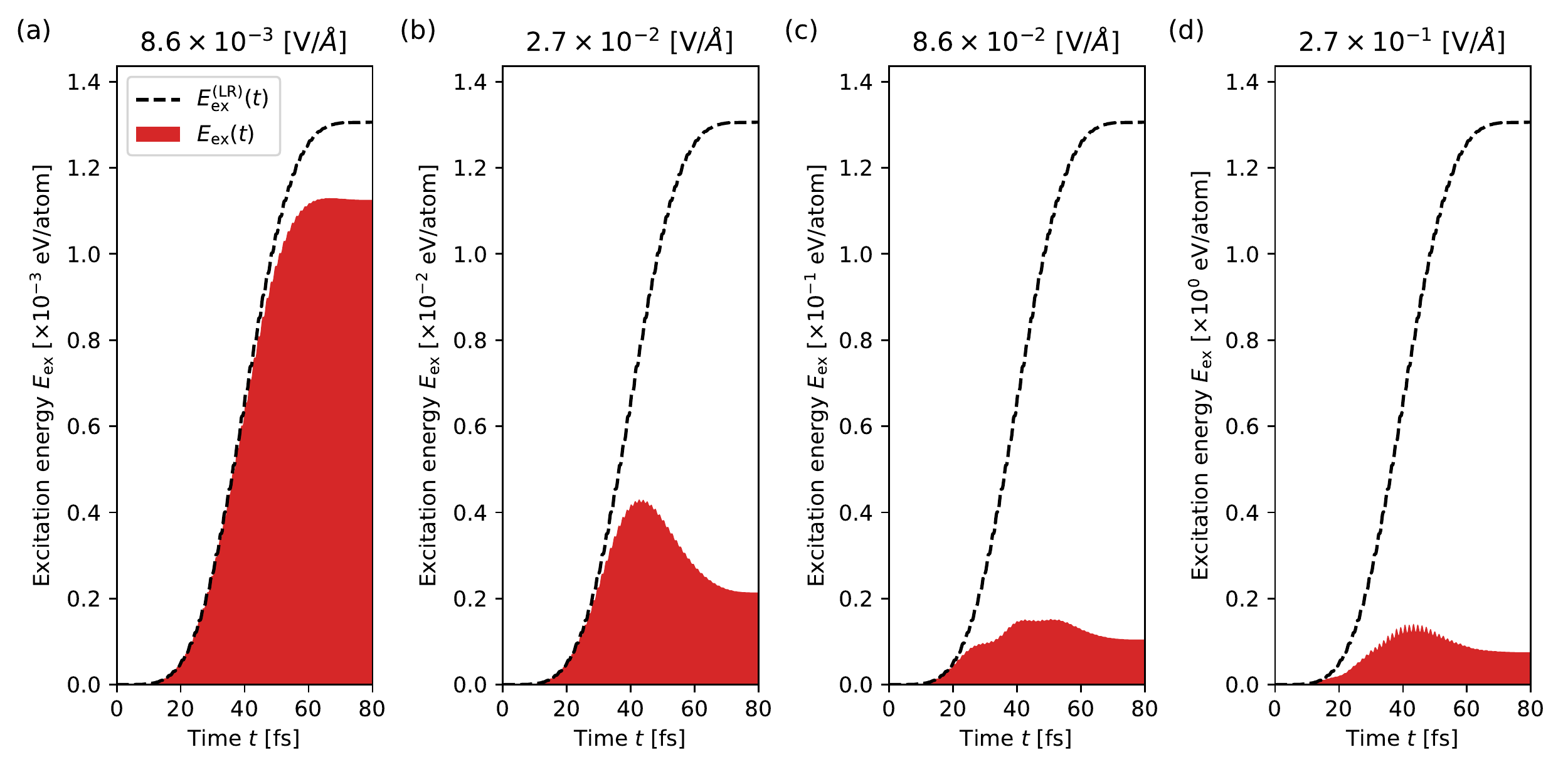}
  }
  \caption{ 
    Excitation energy as a function of time, $E_\mathrm{ex}(t)$, for four cases of different field amplitudes.
    The broken curve is calculated assuming the linear response, $E_\mathrm{ex}^{\mathrm{(LR)}}(t)$.
  }
  \label{fig_energy}       
\end{figure*}

In Fig.~\ref{fig_energy}, we plot $E_\mathrm{ex}(t)$ for four different $E_\mathrm{max}$ amplitudes.
It shows a clear indication of the SA.
In the small amplitude case (a), calculated $E_\mathrm{ex}(t)$ (red) behaves close to the linear response result 
$E^\mathrm{(LR)}_\mathrm{ex}(t)$ (black curve).
As the amplitude increases, $E_\mathrm{ex}(t)$ greatly departs from $E^\mathrm{(LR)}_\mathrm{ex}(t)$.
The ratio of the actual excitation energy to the estimation by the linear response becomes smaller and smaller
as the amplitude increases. 

According to Eq.~(\ref{eq_excitation_linear}), $E_\mathrm{ex}^\mathrm{(LR)}(t)$ increases monotonically with $t$.
However, in Fig.~\ref{fig_energy}(b-d),  $E_\mathrm{ex}(t)$ shows even a descending behavior in the second half of the pulse.
This energy reduction is related to the anti-conducting phase shift shown in Fig.~\ref{fig_phase}.
The presence of the anti-conducting current $ \mathbf{J}(t) \propto - \mathbf{E}(t)$ causes a negative contribution to the 
energy deposition, $W(t) \le 0$.
\begin{figure}
  \resizebox{0.5\textwidth}{!}{%
    \includegraphics{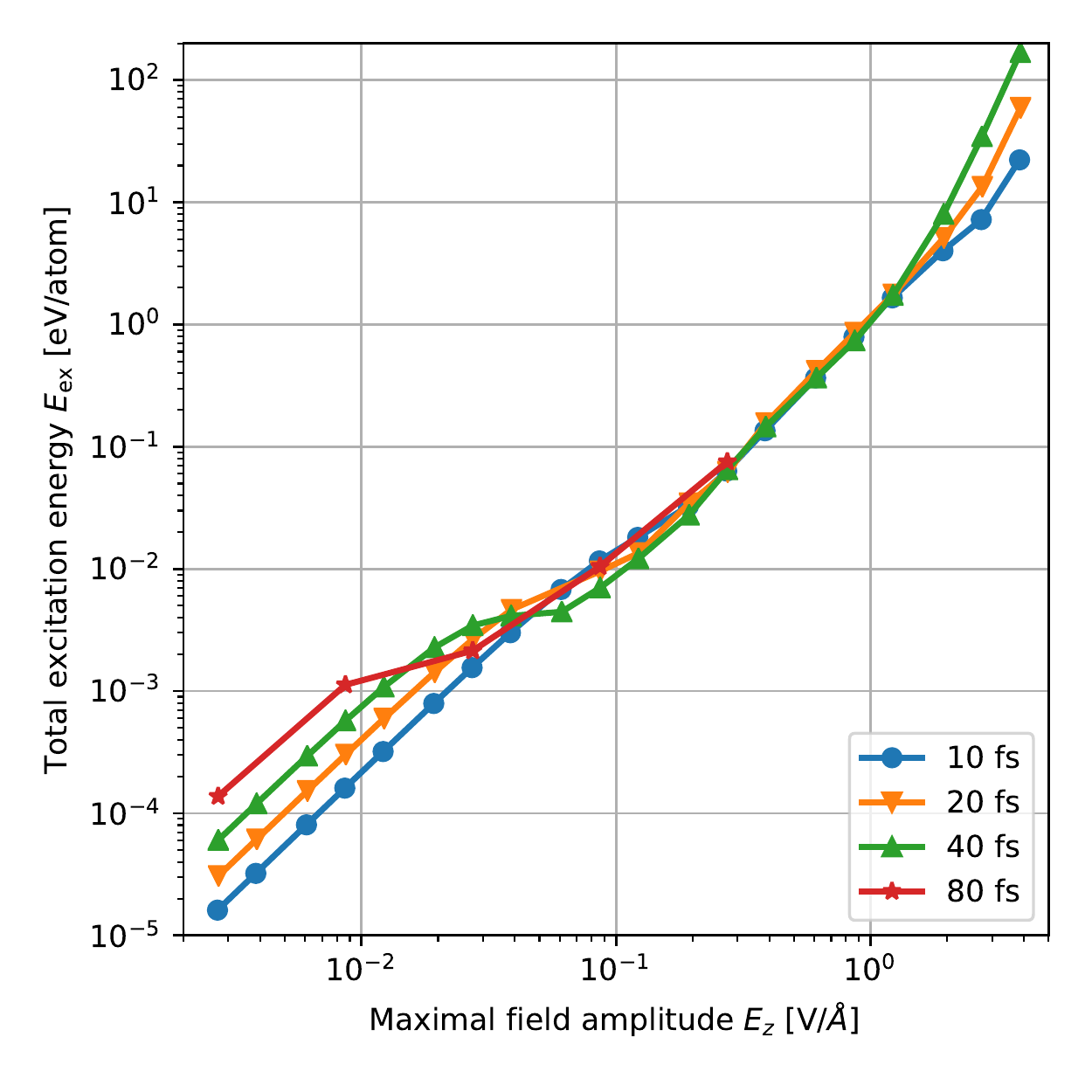}
  }
  \caption{
  Energy transfer from the pulse to the medium as a function of the maximum amplitude of the electric field, $E_\mathrm{max}$,
  for four different pulse durations, $T_P = 10, 20, 40$ and $80~\mathrm{fs}$.
  }
  \label{fig_excitation}       
\end{figure}

In Fig.~\ref{fig_excitation}, we show the total amount of energy transfer from the pulsed electric field 
to electrons in the unit cell that is equal to the electronic excitation energy after the pulse ends, 
$E_\mathrm{ex}(T_P)$.
The energy transfer is plotted against the maximum amplitude of the applied electric field, 
$E_\mathrm{max}$, for four different pulse durations, $T_P=10$~fs, $20$~fs, $40$~fs, and $80$~fs.
This plot again shows a clear indication of the SA.

At weak amplitude $E_\mathrm{max} < 10^{-2}$~V/\AA, the transferred energy is almost quadratic 
in $E_\mathrm{max}$ and is proportional to the pulse duration. 
These behaviors are consistent with the conducting response.
This behavior changes abruptly at and above the amplitude, $E_\mathrm{max} = 10^{-2}$~V/\AA.
The transferred energy shows a smaller slope or even does not increase for a certain region of the 
maximum field amplitude in the region of $10^{-2} \sim 10^{-1}$ ~V/\AA.
At amplitude region of $4 \times 10^{-2} \sim 2 \times 10^0$~V/\AA,
the transferred energy is almost independent of the pulse duration $T_P$. 
This indicates that the saturation takes place at an early stage of the irradiation, 
as quick as 10 fs.
At amplitude region $E_\mathrm{max} > 10^0$ V/\AA, the energy transfer again depends 
on the pulse duration.

\color{red}
\subsection{Comparison with measurements}

Let us compare the maximum filed amplitude that shows the SA with the measured value
of saturation intensity. There are several measurements for single- and multi-layered graphenes.
The measured saturation intensity depends strongly on the duration as well as the frequency
of the laser pulse. In early measurements, longer pulses of more than picosecond duration
was used. Later there has been several measurements that use pulses of femtosecond duration
that correspond well with our caulcations.

In our calculation, the saturation starts at the field amplitude of $4 \times 10^{-2}$~V/\AA~
that corresponds to the intensity of the pulse, $I=1.9 \times 10^{10}$ W/cm$^2$ if we assume
that the maximum field amplitude in the medium is equal to that of the incident pulse.
This should be a good approximation for an extremely thin films.
We compare our value with measurements that are conducted in similar physical conditions
using laser pulses of wavelength of about 800nm.
There are a few measurements using graphite thin films.
Using the pulse of 20 fs duration and for 280 layered graphene, $I_\mathrm{sat} = 3.0 \times 10^{10}$ W/cm$^2$ was reported \cite{Meng2016}.
Using the pulse of 56 fs duration and for 60 layered graphene, $I_\mathrm{sat} = 5.7 \times 10^{10}$ W/cm$^2$ was reported \cite{Norris2012}. 
There are also a few measurements for a single layer of graphene.
Using the pulse of 80 fs duration, $I_\mathrm{sat} = 2.3 \times 10^{10}$W/cm$^2$ was
reported \cite{Kumar2009}. 
Using the pulse of 100 fs duration, $I_\mathrm{sat} = 7.6 \times 10^{11}$W/cm$^2$ was reported \cite{Wang2016ACS}. There was also a measurement using the pulse of duration
200 fs giving $I_\mathrm{sat} = (4 \pm 1) \times 10^{9}$ W/cm$^2$\cite{Xing2010}.
These measurements using light pulses of shorter than a few hundred femtoseconds,
the saturation intensity in our calculation coincides more or less with measurements.
There are also theoretical calculation: $I_\mathrm{sat} = 5 \times 10^{9} \sim
1 \times 10^{10}$ W/cm$^2$ using tight-binding model \cite{Zhang2011}, and
$I_\mathrm{sat} = 6.5 \times 10^{10}$ W/cm$^2$ including various many-body effects \cite{Norris2012}. They also show a reasonable agreement although physical effects
included are different in each approach.

In early measurements, much lower saturation intensity was reported using lower frequency
and longer pulses; for example, $I_\mathrm{sat} = 7.4 \times 10^{6}$ W/cm$^2$ using
a laser pulse of wavelength 1550 nm and duration 3.8 ps \cite{Zhang2012}.
For longer pulses, it becomes important to take account of equilibrium processes
by collisional relaxations.

\color{black}

\begin{figure*}
  \resizebox{1.00\textwidth}{!}{%
    \includegraphics{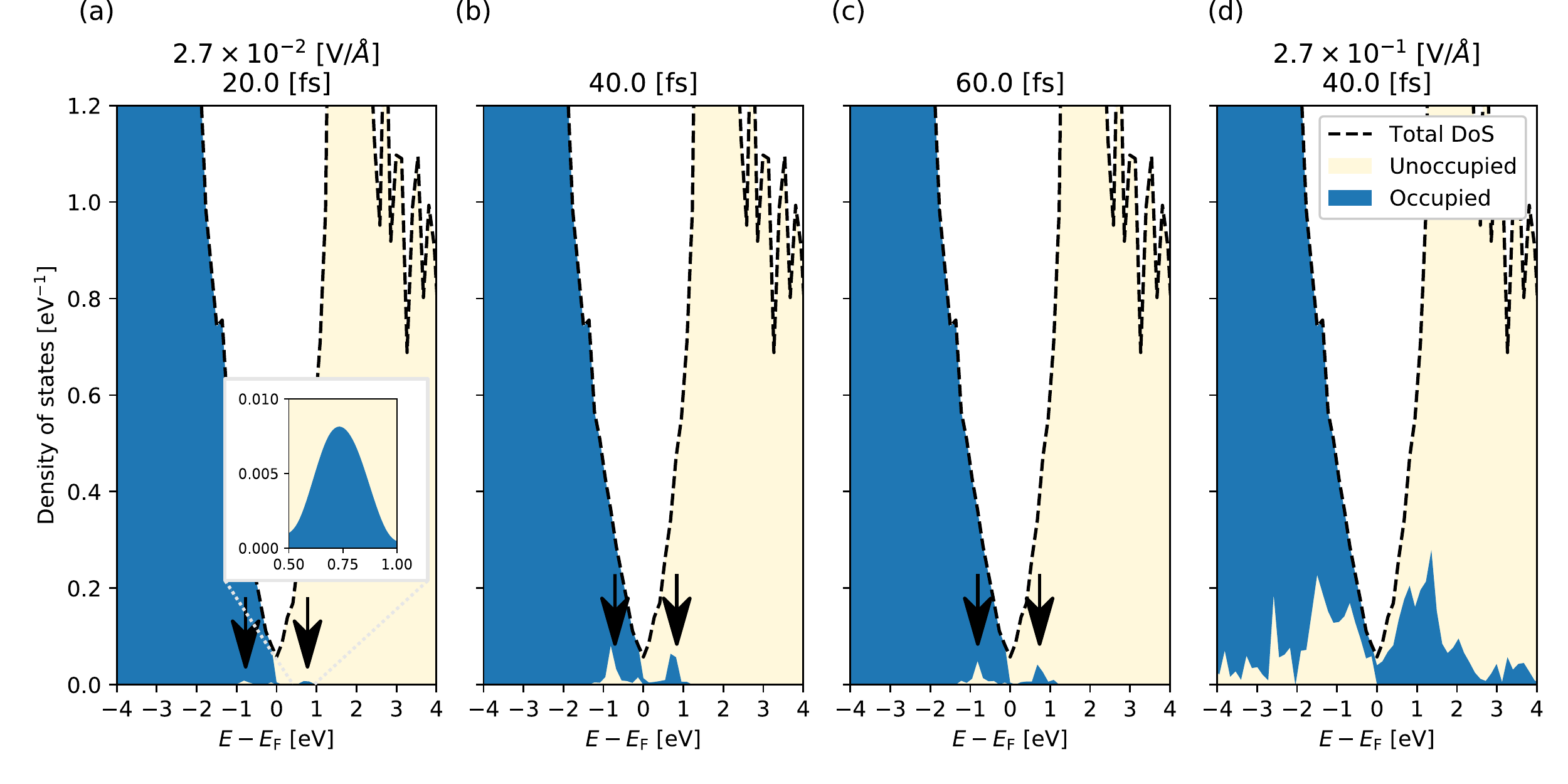}
  }
  \caption{
    Temporal-evolution of the pDoS at the vicinity of the Fermi level. 
    The applied pulse duration and maximal field amplitudes are set as
    $T_P=40~\mathrm{fs}$ and $E_\mathrm{max}=2.7 \times 10^{-2}~\mathrm{V/}$\AA (a-c) and $2.7 \times 10^{-1}~\mathrm{V/}$\AA (d), which correspond to Fig.~\ref{fig_current}(c) and Fig.~\ref{fig_current}(e), respectively.
    The total DoS is plotted as a broken line.
  }
  \label{fig_density_of_states}       
 \end{figure*}

\color{red}
\subsection{Occupation distribution}
\color{black}

Next, we discuss how the occupation of electrons changes during the irradiation of the pulse.
For this purpose, we perform density of states (DoS) analysis of the excited carriers.
We first define the total-DoS (tDoS) $D(E)$ in the ground state by 
\begin{align}
  D(E) 
  =&
  \frac{2}{N_\mathrm{atom}}
  \sum_{b}^\mathrm{(all)} \sum_{\mathbf{k}_i} {
    w_i \delta(E - \epsilon_{b \mathbf{k}_i})
  }
  \label{eq_tdos}
\end{align}
where $\epsilon_{b \mathbf{k}_i}$ is the single particle energy of the orbital $b$ at $\mathbf{k}_i$, and
the summation is taken over all orbitals.
We next define the projected-DoS (pDoS) that indicates the electron occupancy at time $t$,
\begin{align}
  D^{\mathrm{(proj)}}(E)
  =&
  \frac{2}{N_\mathrm{atom}}
   \sum_{b}^\mathrm{(all)}\sum_{\mathbf{k}_i} {
    w_i \delta(E - \epsilon_{b \mathbf{k}_i})
    P_{b \mathbf{k}_i}(t)
  }
  \label{eq_pdos}
\end{align}
with
\begin{align}
  P_{b \mathbf{k}}(t)
  =&
  \sum_{b'}^{\mathrm{(occ)}}
  \left|
    \int_{\Omega_\mathrm{cell}}{
      \mathrm{d} \mathbf{r} \;
      u^\star_{b' \mathbf{k}}(\mathbf{r}; t)
      \;
      u_{b \mathbf{k}+\mathbf{A}(t)/c}(\mathbf{r})
    }
  \right|^2
  \;,
\end{align}
where we use the so-called Houston function, $u_{b \mathbf{k}+\mathbf{A}(t)/c}(\mathbf{r})$ as the reference state
to define the electronic excitation.

 Figure~\ref{fig_density_of_states} shows the tDoS~(\ref{eq_tdos}) and the pDoS~(\ref{eq_pdos}).
 Fig.~\ref{fig_density_of_states}(a)-(c) show the electron occupation at times $t = 20, 40,$ and $60~\mathrm{fs}$
 under the field of maximum amplitude $E_\mathrm{max} = 2.7 \times 10^{-2}$~V/\AA~ that corresponds to 
 Fig.~\ref{fig_current}(c).
 Each timing corresponds to (a) linear, (b) saturation, and (c) anti-conducting responses, respectively. 
 In Fig.~\ref{fig_density_of_states}(a), there appear small peaks of excited electrons and holes at $\approx \pm 0.8$~eV 
 from the Fermi surface. The separation of two peaks is equal to the average frequency of the pulse, $\omega_1=1.55$~eV.
 The maximum density of the excited carriers is two orders of magnitude smaller than the tDoS.
 At $t=40$ fs shown in Fig.~\ref{fig_density_of_states}(b), 
 the excited carrier density is comparable to the tDoS. At this time, the induced current is substantially suppressed
 as seen in Fig.~\ref{fig_current}(c). The occupation change explains the mechanism of the SA: 
 substantial part of the valence electrons are already excited, while the conduction states are mostly filled.
 These two effects suppress the excitation of electrons.
 We note that the occupation change is only a fraction of the tDoS as seen in Fig.~\ref{fig_density_of_states}(b).
 Since electronic excitations take place anisotropically in $k$-space, the saturation appears although only a fraction
 of electrons is excited in energy representation.
 It should, however, noted that the present TDDFT calculation does not take full account of
 the relaxation effects since the $e-e$ collision effects are not included sufficiently.
 As noted in the Introduction, thermalization of the electron distribution within a few tens of femtosecond
 was experimentally reported for graphite \cite{Breusing2009}.
 At $t=60$ fs when the anti-conducting response appears in Fig.~\ref{fig_current}(c), the number of excited carriers 
 becomes small compared with that at the time $t=40$ fs.
 It indicates that the appearance of the anti-conducting current is related with the decrease of the carrier density.
 
 Figure~\ref{fig_density_of_states}(d) shows the occupation distribution at the peak time of the field 
 when an intense field of $2.7 \times 10^{-1}$~V/\AA~ was applied.
 The field amplitude corresponds to that used in Fig.~\ref{fig_current}(e).
 We can see that the carriers distribute in wide energy. It originates from multiple excitation processes
 \color{red}
 in which excited carriers are re-excited by the applied electric field and 
 make transitions to a higher energy band. 
 These final states have a higher DoS than those of Dirac cone and allow existence of high density carriers.
 This high density carriers are expected to contribute to the metallic response
 that was observed in Fig.~\ref{fig_phase}.
 \color{black}
 At this field strength, we still observe a strong SA as seen in Fig.~\ref{fig_energy}(d) 
 and Fig.~\ref{fig_excitation}.

\section{Light propagation}
\label{sec_propagation}

We first summarize a description of 
the light propagation in ordinary electromagnetism. The reflectance $R$ and the penetration 
depth $L_p$ at the surface of the graphite is given by
\begin{align}
  R =
  \frac{
    (n-1)^2 + \kappa^2
  }{
    (n+1)^2 + \kappa^2
  }
  \;,\quad
  L_p =  \frac{c}{2 \omega \kappa}
  \;,
  \label{eq_penetrate_lr}
\end{align}
where $n$ and $\kappa$ are the real and the imaginary parts of the index of refraction of graphite, respectively.
$\omega$ is the frequency of the light.
From the dielectric constant, 
$\epsilon_\mathrm{exp}(\omega_1) \approx 5+9i$ \cite{Taft1965},
the index of refraction of the graphite is given by $n(\omega_1) + i\kappa(\omega_1) = 2.9 + 1.7i$ at  $\omega_1=1.55~\mathrm{eV}$. 
From these values, we obtain $R=0.36$ and $L_p=36~\mathrm{nm}$, respectively.
When a weak light irradiates normally on the bulk graphite surface, the intensity of the pulse decays exponentially as 
$I \propto \exp(-x/L_p)$ at the penetration distance $x$.

\subsection{Thin film of 50 nm}

\begin{figure}
  \resizebox{0.5\textwidth}{!}{%
    \includegraphics{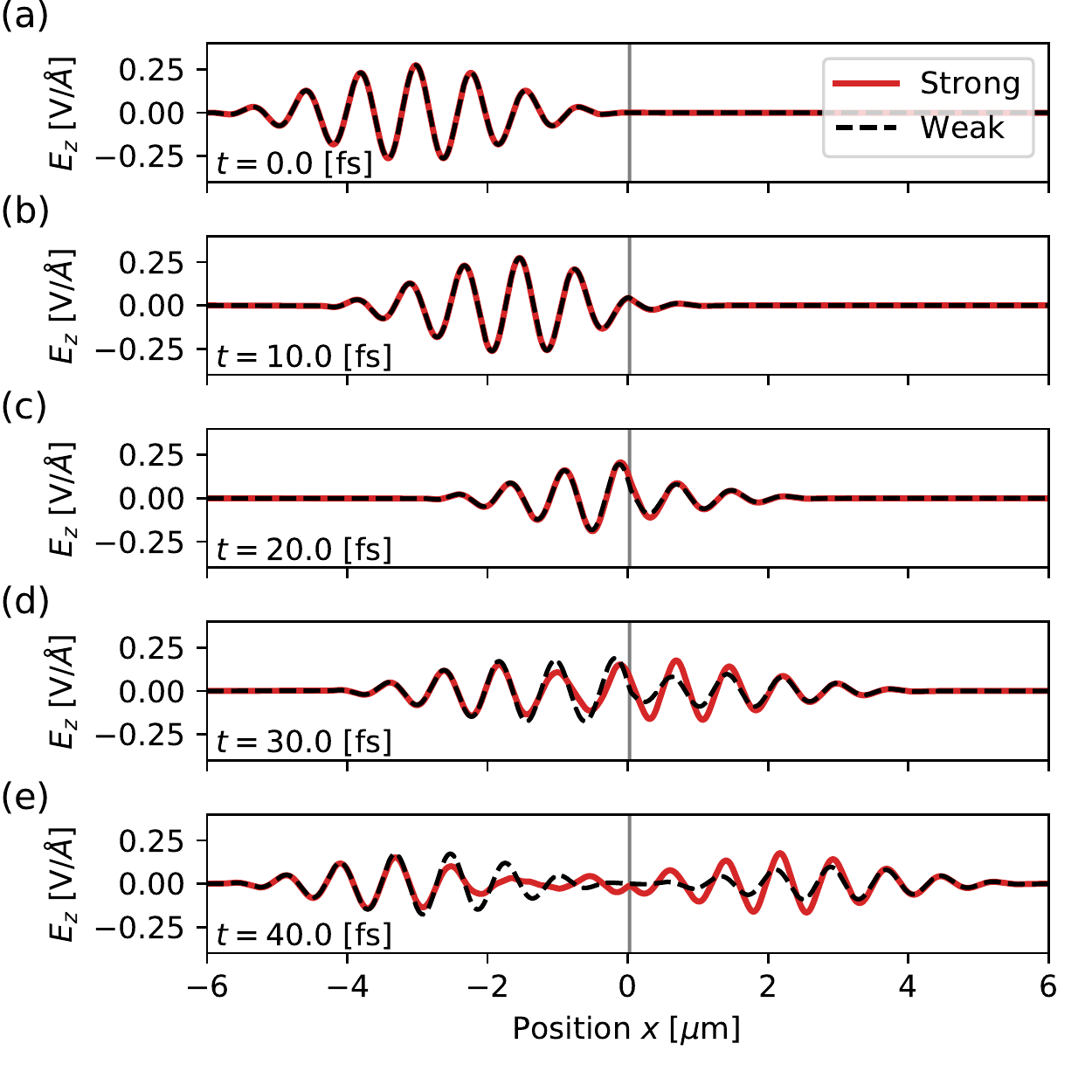}
  }
  \caption{
    Time evolution of the electric field of a laser pulse irradiating normally on the graphite thin film 
    of $50~\mathrm{nm}$ thickness (gray region).
    Two different intensity cases are compared, the maximum intensity in the vacuum of $I=10^{12}~\mathrm{W/cm}^2$ (red) and
    $I=10^{10}~\mathrm{W/cm}^2$ (black). The pulse duration is set as $T_P=20~\mathrm{fs}$.
  }
  \label{fig_propagate_50nm}       
\end{figure}

In this subsection, we will investigate light propagation through a thin film of graphite of 50nm thickness.
This corresponds to about 150 sheets of honeycomb layers.
Since the thickness is comparable to the absorption depth of the graphite, a part of the incident light 
will transmit to the opposite side. In the next subsection, we consider a thicker film without transmission.

Utilizing multiscale method described in Sec.~\ref{sec:multiscale}, 
we calculate light propagation by solving the wave equation of Eq.~(\ref{eq_maxwell1d}) 
combining electron dynamics described by Eq.~(\ref{eq_TDKSmultiscale}).
We consider a linearly polarized light irradiating normally on the graphite thin film.
In Fig.~\ref{fig_propagate_50nm}, we plot the electric field profiles at several times.
The pulsed light initially stays in the left vacuum region as shown in (a), and propagates toward 
positive $x$-direction.
The red line shows the field for the case of a strong pulse with the maximum intensity 
of $10^{12}~\mathrm{W/cm}^2$ at the vacuum. 
The black line shows the field for the case of a weak pulse with the maximum intensity of $10^{10}$ W/cm$^2$.
The black line is multiplied by a factor of 10 so that two curves coincide with each other
in the vacuum region. The difference of two curves shows nonlinear effects in the propagation.
At the final time shown in (e), we clearly observe a consequence of the SA.
Compared with the weak pulse, the transmitted electric field of the strong pulse is much stronger.
Looking in detail, the front part of the two pulses coincides with each other.
After a few cycle, the electric field of the strong pulse is much larger than the weak pulse.
We also find a decrease of the reflected wave for the strong pulse case.

\begin{figure}
  \resizebox{0.5\textwidth}{!}{%
    \includegraphics{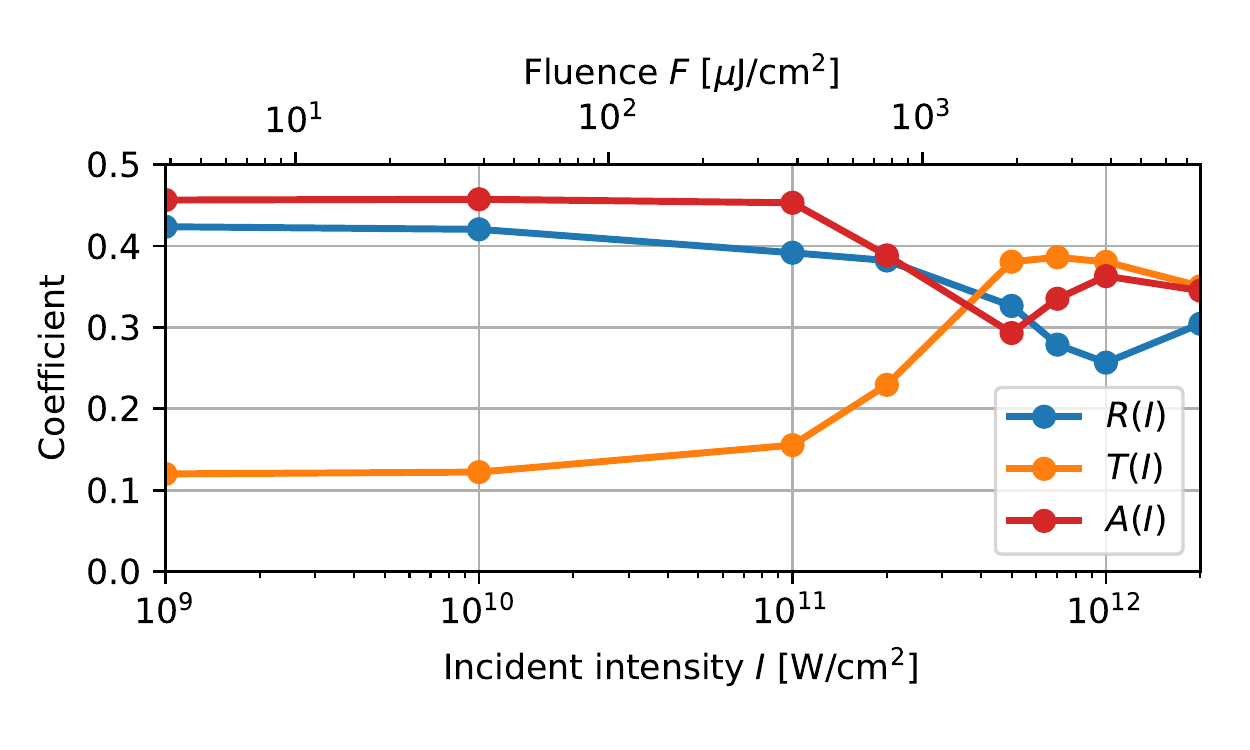}
  }
  \caption{
Intensity-dependence of reflection $R(I)$ and transmittion coefficients $T(I)$ for the thin-film graphite $50~\mathrm{nm}$ thickness.
The pulse duration is set to $T_P = 20~\mathrm{fs}$.
    }
  \label{fig_transmit}       
\end{figure}

In Fig.~\ref{fig_transmit}, we show the intensity(fluence) dependence of the reflectance $R$, the transmittance $T$, 
and the absorbance $A$ that satisfy $R+T+A=1$. The pulse frequency and duration 
are set to 1.55 eV and 20 fs, respectively.
As seen from the figure, nonlinear behavior becomes visible at and above the intensity 
of $1 \times 10^{11}$ W/cm$^2$.
The transmittance increases as high as 0.4, about three times larger than the estimate from the linear response.
Both the reflectance and the absorbance decrease for the strong field.

In Ref. \cite{Meng2016}, there was a measurement of transmission of a laser pulse of 775nm and 20 fs duration
through a multilayer turbostraic graphene of 280 layers. It was reported that the transmission starts to increase at the
intensity of $3 \times 10^{9}$ W/cm$^2$. The increase of the transmission is about 13\% at the intensity
of $3 \times 10^{10}$ W/cm$^2$. 
In Ref. \cite{Norris2012}, there was a measurement of transmission of a laser pulse of 800 nm and 56 fs duration
through 60 layers film. It was reported that the saturation intensity is $5.7 \times 10^{10}$ W/cm$^2$
and the enhancement of 13\% in the transmittance was observed at the intensity of $1.4 \times 10^{11}$ W/cm$^2$.
In view of the difference of the pulse duration and film thickness,
our result is in qualitative agreements with these measurements.

\subsection{Thicker film of 250 nm}

We next consider  the light propagation through a film of 250 nm thickness. 
Since it is much thicker than the penetration depth, we expect a very small transmission.
We consider again a linearly polarized light normally irradiating on the thin film.
The intensity and the duration of the incident pulse is set as $I=10^{12}~\mathrm{W/cm}^2$ 
and $T_P=20~\mathrm{fs}$, respectively.

\begin{figure}
  \resizebox{0.5\textwidth}{!}{%
    \includegraphics{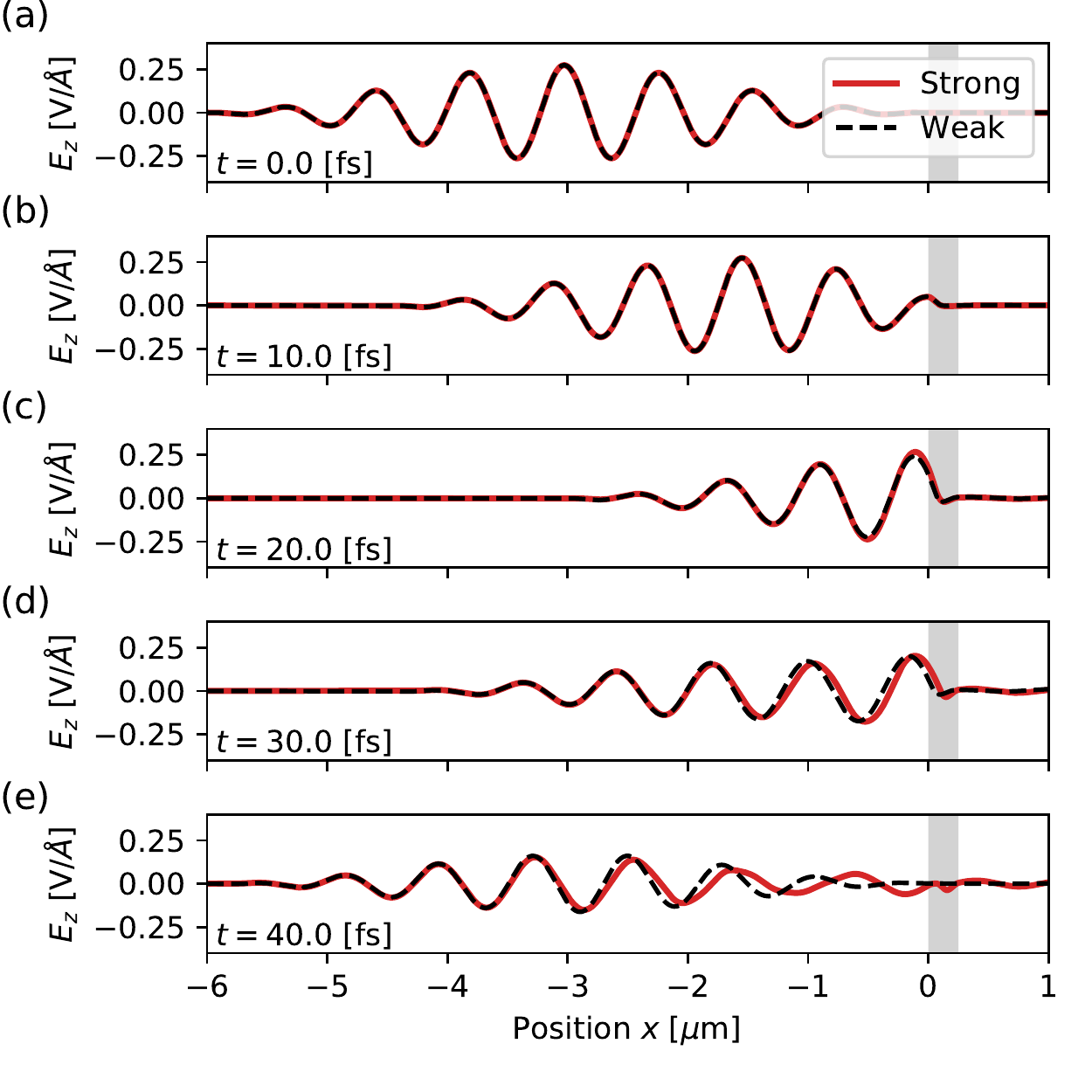}
  }
  \caption{
    Time evolution of a pulsed light irradiating on the graphite thin film of $250~\mathrm{nm}$ thickness (gray region).
    A red line shows the electric field of a strong pulse with maximum intensity of $I=10^{12}$W/cm$^2$. 
    A dashed black line shows the electric field of a weak pulse magnified so that the incident pulses look the same.
  }
  \label{fig_propagate:250nm}       
\end{figure}

We show a typical light propagation in Fig.~\ref{fig_propagate:250nm}.
The red line shows the electric field of the propagating light.
As expected, all the pulse is reflected or absorbed at the film.
The broken black line shows the propagation of the weak pulse,
multiplied with a constant so that it coincides in the linear limit.
Looking at the reflected wave of the bottom panel (e), the front part of the electric field of two intensities
are close to each other.
After $t = 10~\mathrm{fs}$, there appears a phase delay in the strong pulse.

\begin{figure}
  \resizebox{0.5\textwidth}{!}{%
    \includegraphics{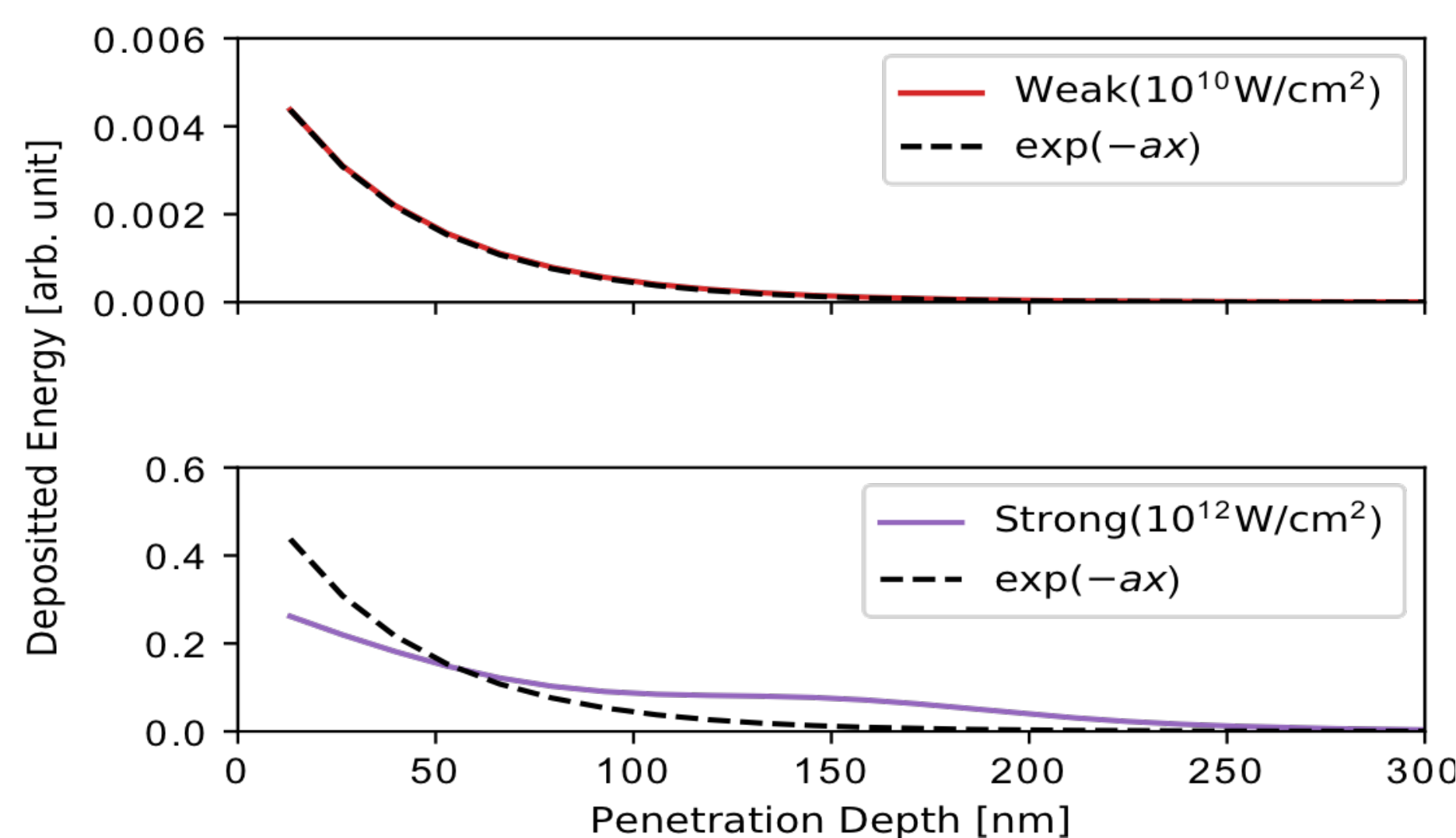}
  }
  \caption{
 Energy deposition as a function of distance from the surface.
 (a) for weak pulse and (b) for strong pulse cases.
     }
  \label{fig_multiscale}       
\end{figure}

In Fig.~\ref{fig_multiscale}, we show the energy deposition from the laser pulse to electrons in the medium
as a function of the distance from the surface, $x$.
For comparison, the energy deposition assuming a linear response is shown by the dashed curve.
It shows the exponential decay as a function of $x$.
In the upper panel, we show the case of the laser pulse of the incident intensity $I=10^{10}$ W/cm$^2$.
In this case, the deposited energy is well fit by the exponential curve, indicating that the absorption can
be described by the ordinary ohmic resistance.
In the lower panel, the absorption of the laser pulse of the intensity $I=10^{12}$ W/cm$^2$ is shown.
At the surface, the absorption is weaker than the estimate by the linear response, which is caused by the SA.
As the distance from the surface increases, the energy deposition becomes larger than the estimate
by the linear response. Because of the SA, the laser pulse is not absorbed efficiently at the surface,
and the pulse can reach deeper inside the materials. It then causes the enhanced energy deposition
deep inside the medium.

\begin{figure*}
  \resizebox{0.95\textwidth}{!}{%
    \includegraphics{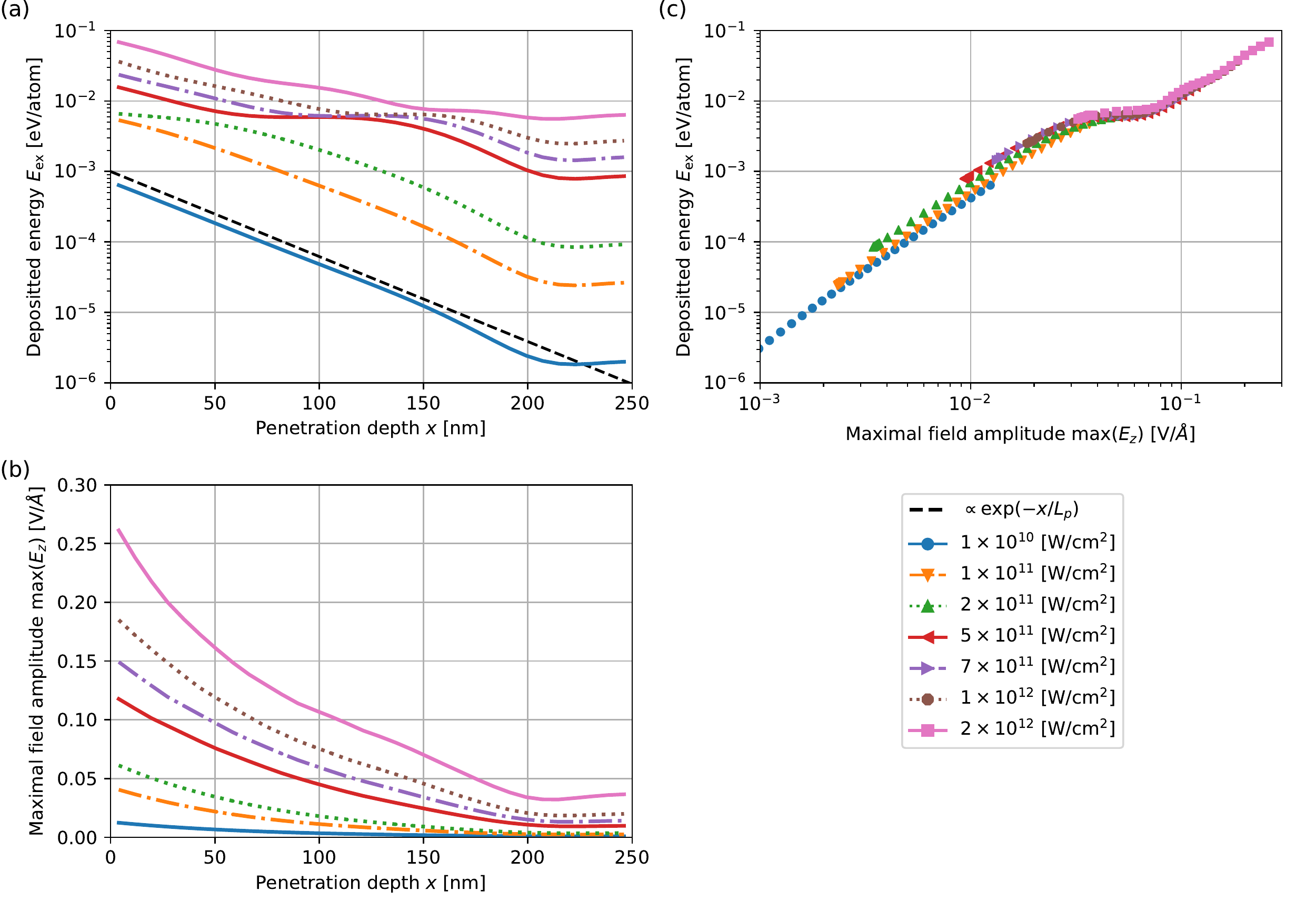}
  }
  \caption{
    (a) Deposited energy $E_\mathrm{ex}(x)$ in the thin-film graphite of 250 nm thickness as a
    function of the distance from the surface $x$, (b) maximum electric field amplitude at the depth $x$,
    and (c) deposited energy plotted against the maximum electric field amplitude.
    The curves shown are  for pulses of intensities $I = 10^9 \sim 10^{13}~\mathrm{W/cm2}$. 
    The pulse duration is set to $T_P = 20~\mathrm{fs}$ for all intensities.
   }
  \label{fig_penetrate}       
  \end{figure*}

To obtain a systematic understanding, we summarize in Fig.~\ref{fig_penetrate}(a) the deposited energy 
$E_\mathrm{ex}(x)$ as a function of the penetration depth from the surface $x$, 
and in Fig.~\ref{fig_penetrate}(b) the maximum electric field at the depth $x$.
They are shown for pulses with different maximum intensities. The pulse duration is chosen to be $T_P=20$ fs.
In Fig.~\ref{fig_penetrate}(a), the black dotted line is an exponential curve 
$\exp(-x / L_P)$ with $L_P = 36$~nm that is expected in the linear response (\ref{eq_penetrate_lr}).

At small intensity $I \le 1 \times 10^{11}$~$\mathrm{W}/\mathrm{cm}^2$, The slope of $E_\mathrm{ex}(t)$ is well fit
by the exponential function, indicating the linear response.
The change in slope around $x = 200 \sim 250$ that can be seen at all intensities 
is due to the reflection at the back-surface ($L = 250$~nm).
At $I=2 \times 10^{11}$~$\mathrm{W}/\mathrm{cm}^2$, the SA becomes appreciable and 
the slope of $E_\mathrm{ex}(x)$ becomes small in the region about $x \le 50$~nm.
From Fig.~\ref{fig_penetrate}(b), the maximum electric field at the surface is about 0.06 V/\AA~
for the incident pulse of $I=2 \times 10^{11}$W/cm$^2$. 
At this field amplitude, a sizable SA is seen in the single cell calculation as seen 
in Fig.~\ref{fig_excitation}.

  In the range of $5 \times 10^{11} \sim 1 \times 10^{12}$~$\mathrm{W}/\mathrm{cm}^2$, 
  a plateau with a very small gradient of $E_\mathrm{ex}(x)$ appears in the vicinity of $x = 100$~nm 
  where the maximum electric field amplitude is close to $4 \sim 7 \times 10^{-2}$~V/\AA.
  This range of amplitude coincides again with the region where the SA is seen in Fig.~\ref{fig_excitation}. 
  The SA makes the pulse penetrate deeper inside the medium than that expected from the linear response.
  For the incident pulse with $I > 5 \times 10^{11}$W/cm$^2$, the energy deposition at the surface 
  increases as the intensity increase,
  and shows a slope that is smaller than that expected from the linear response.
  Here the maximum electric field amplitude exceeds $0.1$~V/\AA~ where, while the SA is still 
  significant, the excitation energy increases as the field amplitude increases as seen 
  in Fig.~\ref{fig_excitation}.

In Fig.~\ref{fig_penetrate}(c), the deposited energy is plotted against the maximum electric field amplitude.
This is constructed from the results of (a) and (b), removing the information of the depth $x$.
We can indeed confirm that the deposited energy is very well correlated with the local value of the
maximum electric field amplitude. 
We find that the deposited energy is almost the same for a region of electric field amplitude,
$0.4$ V/\AA~ $ < E_{\mathrm{max}} < 0.8$ V/\AA. This coincides accurately with the result shown in 
Fig.~\ref{fig_excitation} at $T_P = 20$~fs.

\section{Summary}
\label{sec_summary}

We investigate ultrafast and nonlinear optical responses of graphite thin films
employing the first-principles TDDFT.
First, we investigated optical responses in a unit cell of graphite under a pulsed electric field.
We carried out calculations using a pulsed electric field of various maximum amplitude and duration.
We find that a SA dominates in the nonlinear response.
During the irradiation of a pulsed electric field of a few tens of femtosecond, there appears a
change of optical response from conducting, insulating, and anti-conducting phases.
The SA becomes significant above a certain threshold of the maximum amplitude of the 
applied electric field, around $E_\mathrm{max} \sim 0.01$ V/\AA. 
The threshold amplitude becomes smaller as the pulse duration increases.
The energy transfer from the pulsed electric field to electrons in the medium is found to saturate very quickly,
as fast as 10 fs. 
At sufficiently high amplitude above $E_\mathrm{max} = 1$ V/\AA, the saturation disappears.

We next carried out calculations of a propagation of pulsed light through thin films of graphite.
Coupled multiscale calculations of light propagation and electron motion are carried out,
making it possible to investigate the nonlinear light propagation in the first principles level.
For a thin film of graphite with 50 nm thickness that is comparable to the penetration depth,
the transmitted wave shows a clear indication of the SA.
We investigated the reflectance, transmittance, and absorbance for a pulsed light of 6 fs FWHM duration
and various maximum intensities. We have found the effect of the SA becomes substantial
at and above the maximum intensity of $I = 1 \times 10^{10}$ W/cm$^2$.
The transmittance increases from 0.12 in the linear response region to 0.4.
We also performed a calculation for a thick sample of 250 nm thickness where transmitted wave is very small.
We find that the intense pulse penetrates deeper inside the medium by the SA.
Looking in detail the energy deposition from the light pulse to electrons, there appears a plateau region
of absorbance for the field amplitude of $0.04 < E_\mathrm{max} < 0.07$ V/\AA.
This makes the penetration of the light pulse deeper inside the medium for pulses of maximum intensity
$I > 2 \times 10^{11}$ W/cm$^2$.

\section{Acknowledgements}
We acknowledge the support by MEXT as a priority issue theme 7 to be tackled by using Post-K Computer 
and JST- CREST under Grant No. JP-MJCR16N5 and by JSPS KAKENHI under Grant Nos. 20H02649 and 20K15194. 
M.U. also acknowledge Iketani Science and Technology Foundation.
Calculations are carried out at Oakforest-PACS at JCAHPC through the Multidisciplinary Cooperative Research Program in CCS, 
University of Tsukuba, and through the HPCI System Research Project (Project Nos. hp180088 and hp190106).

\bibliography{refs}

\end{document}